\documentclass[aps,showpacs,twocolumn,prl]{revtex4}
\usepackage{graphicx}
\usepackage{color} 

\usepackage{amsfonts}

\usepackage{amssymb}

\begin{document}

\newcommand{\pderiv}[2]{\frac{\partial #1}{\partial #2}}
\newcommand{\deriv}[2]{\frac{d #1}{d #2}}
\newcommand{\eq}[1]{Eq.~(\ref{#1})}  
\newcommand{\infint}{\int \limits_{-\infty}^{\infty}}

\title{Dissipative Effects in Nonlinear Klein-Gordon Dynamics}

\vskip \baselineskip

\author{A.R. Plastino$^{1,2}$}\thanks{arplastino@unnoba.edu.ar}
\author{C. Tsallis$^{2,3,4}$}

\address{
$^{1}$CeBio y Secretar\'{\i}a de Investigaci\'on, Universidad Nacional Buenos Aires-Noreoeste
(UNNOBA) and Conicet, Roque Saenz Pe\~na 456, Junin, Argentina \\
$^{2}$Centro Brasileiro de Pesquisas F\'{\i}sicas, Rua Xavier Sigaud 150,
22290-180  Rio de Janeiro-RJ, Brazil \\
$^{3}$National Institute of Science and Technology for Complex Systems,
Rua Xavier Sigaud 150, 22290-180, Rio de Janeiro - RJ, Brazil \\
$^{4}$Santa Fe Institute, 1399 Hyde Park Road, Santa Fe, New Mexico 87501,
USA}

\date{\today}

\newpage

\begin{abstract}
We consider dissipation in a recently proposed nonlinear
Klein-Gordon dynamics that admits  soliton-like solutions of the power-law
form $e_q^{i(kx-wt)}$, involving the $q$-exponential function
naturally arising within the nonextensive thermostatistics [$e_q^z \equiv [1+(1-q)z]^{1/(1-q)}$,
with $e_1^z=e^z$]. These basic solutions behave like free particles,
complying, for all values of $q$, with the de Broglie-Einstein relations $p=\hbar k$,
$E=\hbar \omega$ and satisfying a dispersion law corresponding to the
relativistic energy-momentum relation $E^2 = c^2p^2  + m^2c^4 $. The
dissipative effects explored here are described by an evolution equation
that can be regarded as a nonlinear version of the
celebrated telegraphists equation, unifying within one single theoretical framework
the nonlinear Klein-Gordon equation,  a nonlinear Schroedinger
equation, and the power-law diffusion (porous media) equation.
The associated dynamics  exhibits physically appealing
soliton-like traveling solutions
of the $q$-plane wave form with a complex frequency $\omega$
and a $q$-Gaussian square modulus profile.

\vskip \baselineskip

\end{abstract}
\maketitle


The spatio-temporal behavior of a wide family of physical systems
 and processes is described by non-linear partial differential equations
  \cite{S07,PZ04,SS99}.  This has stimulated  an increasing research
 activity on the  dynamics associated  with a class of evolution  equations
 that includes nonlinear versions of the Schr\"odinger~\cite{SS99,NMT2011,PSNT2014}
and the Fokker-Planck~\cite{F05,CearaRio2010,Mauricio2012,plastino95,FF05} ones.
Our  main concern here will be with a  family of  telegraphists-like equations
describing dissipative effects in the context of a recently advanced nonlinear Klein-Gordon
dynamics (NLKGD) \cite{NMT2011} related to  nonextensive statistical mechanics
and the associated nonadditive entropies \cite{T88,TBook2009,B09}.

The free-particle nonlinear Klein-Gordon equation proposed in  \cite{NMT2011}
 has a nonlinearity in the mass term which, in contrast to what happens
in the standard linear case, is proportional to a power of the wave function
$\Phi(x,t)$.  The salient feature of the  NLKGD  introduced in \cite{NMT2011}
is that it exhibits  soliton-like localized solutions where the space-time dependence
of the wave function $\Phi(x,t)$ occurs solely  through the combination  $x-vt$.
Consequently,  one has a space translation at a constant velocity $v$
without change in the wave function's shape.
These soliton-like solutions are known as $q$-plane waves
and are compatible, for all values of $q$, with the Planck and de Broglie relations,
satisfying $E=\hbar w$ and $p=\hbar k$, with $E^2=  c^2 p^2 + m^2 c^4$.
It was shown in \cite{NMT2011} that there is also a nonlinear Schroedinger equation
(often referred to as the NRT Schroedinger equation) with
a nonlinearity in the Laplacian term, that also admits $q$-plane wave solutions,
which are compatible with the non relativistic relation $E = p^2/2m$. Under Galilean
transformations the $q$-plane wave solutions of the NRT Schroedinger equation
recover the transformations rules of the linear Schr\"{o}dinger
equation \cite{PT2013}. The NRT equation satisfied by the $q$-plane waves can be
obtained from a field theory based upon an action variational
principle \cite{NMT2012}. These properties suggest that the $q$-plane wave solutions of both the
nonlinear NRT Schroedinger and Klein-Gordon equations studied in
\cite{NMT2011} can be regarded
as a new field theoretical description of particle dynamics
that may be relevant in diverse areas of physics, including nonlinear optics,
superconductivity, plasma physics, and dark matter \cite{galgani,NMT2012}.

As already mentioned, the dissipative nonlinear Klein-Gordon dynamics that we are going
to explore here is described by a family of telegraphists-like equations.
The standard telegraphists equation constitutes a cornerstone of mathematical physics,
with deep theoretical significance and manyfold applications
\cite{Z89,K74,GJKS84,BL2005,RMV2004,HG2015,LVD98,LS2013,EL2014}.
Historically the telegraphists equation
was first formulated to describe leaky electrical transmission lines.
In one dimension it has the form,

\begin{equation}
\frac{1}{c^2} \frac{\partial^2 \Psi}{\partial t^2} - \frac{\partial^2 \Psi}{\partial x^2}
+ \delta \frac{\partial \Psi}{\partial t} = 0.
\end{equation}

\noindent
This equation corresponds to phenomena intermediate between wave propagation and diffusion.
It can also be regarded as a wave equation with a damping effect described by the term having
the first time derivative. It admits a statistical interpretation in terms of a Poisson process
(dichotomous diffusion) associated with particles that move with constant speed and change the direction of
motion at random times \cite{K74,GJKS84,BL2005}. It has profound (and surprising) connections
with quantum mechanics, being intimately related to the Dirac equation \cite{GJKS84}.
The applications of the telegraphists equation are diverse. We can mention correlated
random walks \cite{Z89}, tunneling processes \cite{RMV2004}, diffusion phenomena in optics \cite{HG2015,LVD98},
and cosmic ray transport \cite{LS2013,EL2014}.

The  $q$-plane waves arise naturally within a
theoretical  framework  where the Boltzmann-Gibbs (BG)
entropy and statistical mechanics are generalized through
the introduction of a power-law entropic functional $S_q$
characterized by an index $q$
(BG being recovered in the limit $q \rightarrow 1$).
Recent progress along these lines of enquiry includes,
for instance, nonlinear extensions of various important equations
of physics and new forms of the Central Limit Theorem~\cite{CLT}.
The $q$-Gaussian distributions, which generalize
the standard Gaussian distribution and  arise from  the optimization of
the $q$-entropy~\cite{T88}, or as solutions of the corresponding nonlinear
Fokker-Planck equation~\cite{plastino95}, play a central role
within these developments.  They have
found several interesting applications to the analysis of
recent experimental findings \cite{TBook2009}. These applications
concern diverse physical systems including, among others,
(i) cold atoms in dissipative optical
lattices~\cite{douglas06};
(ii) quasi-two dimensional dusty
plasmas ~\cite{liugoreeprl08};
(iii)  ions in radio frequency traps interacting with a buffer gas~\cite{devoe};
(iv) RKKY spin glasses, like CuMn and
AuFe~\cite{pickup};
(v) Overdamped motion of vortices in type II superconductors~\cite{soaresevaldo}. More generally, $q$-exponential distributions
have also been applied to a variegated set of physical scenarios.
As recent examples we can mention the description of the transverse momentum
spectra in high-energy proton-proton and proton-antiproton collisions
\cite{WWCT2015}, universal financial \cite{Ludescher} and biological \cite{Bunde} laws, among others.

Of the three nonlinear dynamical equations
admitting $q$-plane wave solutions advanced in \cite{NMT2011}
(the NRT Schroedinger, and the $q$-nonlinear Klein-Gordon and Dirac equations)
the NRT Schroedinger equation is the one that has been more intensively studied so far.
Recent advances along these lines are the investigation of an associated
field theory \cite{NMT2012}, of the effects of Galilean transformations \cite{PT2013},
of quasi-stationary, wave packet, and uniformly accelerated solutions \cite{PSNT2014,qusolutions},
and of its relation with the Bohmian formulation of quantum mechanics \cite{PPP2014}.

The nonlinear Klein-Gordon dynamics introduced in \cite{NMT2011}
is governed by the field equation,

\begin{equation}
\label{NOLIKGE}
 \frac{1}{c^2}{\partial^2 \over \partial t^2}
\left[ \frac{\Phi(\vec{x},t)}{\Phi_{0}} \right]
-
\nabla^{2} \left[ \frac{\Phi(\vec{x},t)}{\Phi_{0}}  \right]
+ q \frac{m^2 c^2}{\hbar^2} \left[ \frac{\Phi(\vec{x},t)}{\Phi_{0}} \right]^{2q-1}
= 0,
\end{equation}

\noindent
where ${\vec x}\in  {\mathbb{R}}^d $,
$\nabla = \left(\frac{\partial}{\partial x_1}, \ldots, \frac{\partial}{\partial x_d} \right)$
is the $d$-dimensional $\nabla$-operator,
 $q\ge 1$ and the real, positive constant  $\Phi_{0}$
leads to correct physical dimensionalities
for all terms (this scaling becomes irrelevant only in the limit
case of  the linear Klein-Gordon equation, that is, for $q=1$). The constant $\Phi_{0}$
constitutes a parameter characterizing the evolution equation
(\ref{NOLIKGE}) itself (that is, it should not be regarded as part of the initial conditions).
The dynamical equation (\ref{NOLIKGE}) can be obtained within
 a classical field theory derived from an appropriate Lagrangian variational
 principle \cite{RN2013}.

The $q$-plane wave solutions of the field equation ({\ref{NOLIKGE})
are given by a $q$-exponential function evaluated on a pure 
imaginary argument, which corresponds to the principal value
of
\begin{equation}
\label{eq:compqexp}
\exp_{q}(iu) = \left[ 1 + (1-q)iu \, \right]^{\frac{1}{1-q}};
\,\exp_{1}(iu) \equiv \exp(iu),
\end{equation}
where $u \in {\mathbb R}$.
The basic relations satisfied by the above function are ~\cite{B98},
\begin{eqnarray}
\label{eq:propcompqexp1}
\exp_{q}(\pm iu) &=& \cos_{q}(u) \pm i \sin_{q}(u)~, \cr
\cos_{q}(u) &=& r_{q}(u)
\cos \left\{ {1 \over q-1} {\rm arctan}[(q-1)u] \right\}~, \cr
\sin_{q}(u) &=& r_{q}(u)
\sin \left\{ {1 \over q-1} {\rm arctan}[(q-1)u] \right\}~, \cr
r_{q}(u) &=& \left[1+(1-q)^{2}u^{2} \right]^{1/[2(1-q)]}~,
\end{eqnarray}
and
\begin{eqnarray}
\label{eq:propcompqexp5}
\exp_{q}(iu)\exp_{q}(-iu) \! & = &
\! [r_{q}(u)]^2  \! = \! \exp_q(-(q-1)u^2), \cr
\exp_{q} (i u_{1})  \exp_{q} (i u_{2})
\! & \ne & \! \exp_{q} \left[ i(u_{1} + u_{2}) \right], \,\, (q\ne1)~.
\end{eqnarray}
It is plain from Eqs.~(\ref{eq:propcompqexp1})-(\ref{eq:propcompqexp5})
that a $q$-exponential with a pure imaginary argument, $\exp_{q}(iu)$,
exhibits an oscillatory behavior with a $u$-dependent amplitude $r_{q}(u)$.
It can immediately be verified that the function $\exp_{q}(iu)$ is
of square integrable for $1<q<3$, the concomitant
integral being divergent both for $q \le 1$ and $q \ge 3$.

The $d$-dimensional $q$-plane wave solution of equations (\ref{NOLIKGE})
is given by

\begin{equation}
\label{eq:3dqsolwaveeq}
\Phi(\vec{x},t) =  \Phi_{0} \, \exp_{q} \left[ i (\vec{k} \cdot \vec{x}
-\omega t) \right],
\end{equation}

\noindent
If we take into account that $d\exp_q(z)/dz=[\exp_q(z)]^q$ and
$d^2\exp_q(z)/dz^2=q[\exp_q(z)]^{2q-1}$ we obtain, for the $(d+1)$-dimensional
d'Alembertian operator,

\begin{equation}
\label{dalambertian}
\left[  \frac{1}{c^2} \frac{\partial^2}{\partial t^2}
 - \nabla^{2}  \right]   \left(\frac{\Phi}{\Phi_0} \right)   =
- q  \left[  \left( \frac{\omega}{c}\right) ^{2}  -
\left( \sum_{n=1}^{d} k_{n}^{2} \right)
\right]
\left( \frac{\Phi}{\Phi_0} \right)^{2q - 1}.
\end{equation}

\noindent
Using the above relation it can be verified that the $q$-plane wave ansatz
(\ref{eq:3dqsolwaveeq}) satisfies the nonlinear field equations (\ref{NOLIKGE})
if the  frequency $\omega$ and the momentum $k$ comply with the relation,

\begin{equation}
\omega^2 = c^2 k^2 + \frac{m^2 c^4}{\hbar^2 }.
\end{equation}

Making now, through
the celebrated de Broglie and Planck relations, the identifications
$\vec{k} \rightarrow \vec{p}/\hbar$ and $\omega \rightarrow E/\hbar$,
it is plain that  the $q$-plane waves are solutions of equation
(\ref{NOLIKGE}) satisfying  $E^2  = c^2p^2 + m^2c^4 $.  That is, they comply with
 the energy spectrum of a relativistic free particle for all values of $q$}.
Therefore (\ref{NOLIKGE}), together with its solution
\eq{eq:3dqsolwaveeq}, constitute promising candidates for describing
interesting types of physical phenomena.


The structure of the nonlinear Klein-Gordon equation in $d$-dimensional space
 \cite{NMT2011} comprises two parts:
a term corresponding to the linear $(d+1)$-dimensional wave equation plus a nonlinear mass
term proportional to a power of the wave function.  Here we are going to introduce a family of evolution equations
endowed with a more general power-law nonlinear term (that incorporates the one appearing in the nonlinear
Klein-Gordon as a particular instance) that preserve the soliton-like $q$-plane wave solutions
of the NLKGD.

Let us consider the equation of motion,

\begin{eqnarray} \label{telequ}
&& \frac{1}{c^2} \frac{\partial^2 } {\partial t^2} \left[ \frac{\Phi(\vec{x},t)}{\Phi_{0}}  \right]
 -  \nabla^2 \left[ \frac{\Phi(\vec{x},t)}{\Phi_{0}}  \right]  + \cr
 &q& \sum_{i=1}^L \delta_i  \left[ \frac{\Phi(\vec{x},t)}{\Phi_{0}}  \right]^{\alpha_i^{(1)}}
\left( \frac{\partial }{\partial t} \left[ \frac{\Phi(\vec{x},t)}{\Phi_{0}}  \right] \right)^{\alpha_i^{(2)}}  =  0,
\end{eqnarray}

\noindent
characterized by the $(3L+1)$ parameters $q$,
$\delta_i$,  $\alpha_i^{(1)}$, and $\alpha_i^{(2)}$, with $i = 1, \ldots, L$.
As in the case of the nonlinear Klein-Gordon equation, the constant
$\Phi_0$ guaranties the correct dimensionalities of the different
terms appearing in (\ref{telequ}). The parameters $q$,  $\alpha_i^{(1)}$, and $\alpha_i^{(2)}$ are
dimensionless, while the dimensions of the $\delta_i$'s depend on the values
of the exponents  $\alpha_i^{(1)}$, and $\alpha_i^{(2)}$.
It can be verified after some algebra that the evolution equation
(\ref{telequ}) admits solutions of the $q$-plane wave form
(from now on we adopt the notation $\Psi = \Phi/\Phi_0$),

\begin{equation} \label{wavequ}
\Psi = [  1 + (1- q) i (\vec{k} \cdot \vec{x}
-\omega t )  ]^{\frac{1}{1-q}},
\end{equation}

\noindent
with $q>1$,  provided that the exponents  $\alpha_i^{(1)}$, and $\alpha_i^{(2)}$
comply with the consistency relation,

\begin{equation} \label{alfaqu}
\alpha_i^{(1)} +  q  \alpha_i^{(2)}  = 2q-1  ,  \,\,\,\,\,\,\,\,  i=1,\ldots, L.
\end{equation}

\noindent
and the wave number vector $ \vec{k}  $ is related to the frequency $\omega$
through the dispersion relation,

\begin{equation} \label{dispequ}
-\frac{\omega^2}{c^2} + k^2 + \sum_{i=1}^L \delta_i (-i \omega)^{\alpha_i^{(2)}} = 0,
\end{equation}

\noindent
where $k^2 = \vec{k} \cdot \vec{k}$. The $q$-plane wave (\ref{wavequ})
constitutes a solution of (\ref{telequ}) for any $q$. However, we shall consider only
$q>1$, yielding $\lim_{|{\vec k} \cdot {\vec x}|\to \infty} |\Psi|^2  = 0$,
while for $q<1$ one has $\lim_{|{\vec k} \cdot{\vec x}| \to \infty} |\Psi|^2  = \infty$.
In the case of $L=1$, $\delta_1 = m^2 c^2/\hbar^2$, and $\alpha_1^{(2)} = 0$
equation (\ref{telequ})  coincides with the nonlinear Klein-Gordon equation proposed in
\cite{NMT2011} which, in turn,  reduces to the standard Klein-Gordon equation in the limit $q \to 1$.
On the other hand, for $L=1$, $\delta_1 > 0$,  $\alpha_1^{(2)} = 1$, and $q\to 1$
the standard linear telegraphists equation is recovered. Other relevant equations
are obtained as particular limit cases of  (\ref{telequ}). For instance,  the limit
$c \to \infty$  corresponds to equations respectively equivalent to the NRT nonlinear Schroedinger
equation (for $L=1$, $\delta_1$ pure imaginary, and $\alpha_1^{(2)} = 1$)
and to the porous media equation  (for $L=1$,  $\delta_1$ real, $q\ne 1 $,
and $\alpha_1^{(2)} = 1$) .

We shall assume a wave vector $ \vec{k}  $ with real components (this choice
can be regarded as a choice determining the form of the initial form of the
wave function at $t=0$). The frequency of the $q$-plane wave solutions is  then
determined by solving the  dispersion relation (\ref{dispequ}) for $\omega$.
In general we are going to have a complex frequency,

\begin{equation}
\omega = \omega_a + i \omega_b,
\end{equation}

\noindent
the imaginary part $\omega_b$ corresponding to the dissipation
effects implied by the evolution equation  (\ref{telequ}).

The qualitative features of the dynamics associated with the $q$-plane
wave solutions can be clarified by considering the behavior of its
squared wave function profile. We have,

\begin{equation} \label{moduqu}
\left|  \Psi  \right|^2 = A \left[ 1 - (1-q)
(\vec{x}-  \vec{x}_0)^{T}   {\cal B}  (\vec{x}-  \vec{x}_0)
\right]^{\frac{1}{1-q}},
\end{equation}

\noindent
where

\begin{equation} \label{ampliqu}
A  = \left[ 1   + (1-q) \omega_b  t  \right]^{\frac{2}{1-q}} \equiv (e_q^{\omega_bt})^2,
\end{equation}

\begin{equation} \label{centroqu}
\vec{x}_0  = \frac{\omega_a t}{k^2} \vec{k},
\end{equation}

\noindent
and ${\cal B}$ is an $L \times L$ matrix with elements

\begin{equation} \label{betaqu}
\beta_{ij} = \frac{(q-1) k_i k_j }{ \left[ 1 + (1-q)\omega_b t  \right]^2} \,.
\end{equation}

\noindent
In (\ref{moduqu}), following a standard notational convention, $(\vec{x}-  \vec{x}_0)$
is to be understood as a column vector, while $(\vec{x}-  \vec{x}_0)^{T} $
stands for the concomitant row vector.

We see that the squared modulus profile $|\Psi|^2$ has the shape of a multi-valuated $q$-Gaussian
with both its center $\vec{x}_0$ and amplitude $A$ being time dependent. The center  $\vec{x}_0$
(where  $|\Psi|^2$ adopts its maximum value $A$) moves uniformly with a constant
velocity $d\vec{x}_0/dt = \vec{k} \omega_a/k^2$. In the dissipative case, corresponding to $\omega_b < 0$
(remember that we are considering $q> 1$),
the amplitude $A$ decreases according to a $q$-exponential law. That is, we have,
$A = \exp_q^2(\omega_b t)$. Equations (\ref{wavequ}) and (\ref{moduqu}) are written with respect to an arbitrary
cartesian spatial reference frame. If we choose a reference frame oriented in such a way that the $x_1$
axis points in the direction of the wave vector $\vec{k}$,  then $\Psi$ becomes independent of the
remaining $(d-1)$ spatial coordinates, and $|\Psi|^2$ can be expressed in terms of one single
coordinate $x=x_1$,

 \begin{equation} \label{moduqunde}
\left|  \Psi  \right|^2 = A \left[ 1 - (1-q)
\beta(x - x_0)^2 \right]^{\frac{1}{1-q}}   \equiv Ae_q^{-\beta (x-x_0)^2},
\end{equation}

\noindent
with $A$ given by (\ref{ampliqu}) and

\begin{equation}
\beta =   \frac{(q-1) k^2}{ \left[ 1 + (1-q)\omega_b t  \right]^2}.
 \end{equation}

\noindent
We see that in the dissipative case we have $\beta \to 0$
when  $t \to \infty$. That is, the $q$-plane wave solution
becomes less localized as it evolves.  From now one, we are
going to work with a reference frame oriented as explained above,
so that we are going to consider an effective one dimensional problem

An interesting special case is given by $L=2$,
$\alpha_1^{(1)}= 2q-1 $,
 $\alpha_1^{(2)} = 0 $, $\delta_1 =  m^2 c^2/\hbar^2$
$\alpha_2^{(1)} = -1 $,
$\alpha_2^{(2)} = 2 $, and $\delta_2  = \delta >0$.
This case yields a non-dissipative,
time reversible dynamics. The dispersion relation is,

\begin{equation} \label{nondisidisp}
 - \left(  \frac{1}{c^2}  + \delta    \right)  \omega^2  + k^2  + \frac{m^2 c^2}{\hbar^2}  = 0.
\end{equation}

\noindent
This dispersion relation is consistent with the relativistic energy-momentum
relation with an effective velocity of light,

\begin{equation}
c^* = \frac{c}{\sqrt{1 + \delta c^2}} < c,
\end{equation}

\noindent
and an effective mass,

\begin{equation}
m^* = m \sqrt{1 + \delta c^2} \ge m.
\end{equation}

\noindent
For zero rest mass $(m=0)$ we obtain
 the evolution equation,

\begin{equation} \label{qoncer}
\frac{1}{c^2} \frac{\partial^2 \Psi}{\partial t^2} - \frac{\partial^2 \Psi}{\partial x^2}
+ {\bar \delta}  \frac{1}{\Psi}  \left( \frac{\partial \Psi}{\partial t} \right)^2 = 0,
\end{equation}

\noindent
where, ${\bar \delta} \equiv q \delta$. This notation stresses the fact that the
structure of the above equation is $q$-independent. This equation has the
remarkable property of admitting $q$-plane wave soliton-like solutions that propagate
without changing shape and with constant velocity, {\it for all values $q>1$}.
In contrast to what happens with the standard linear wave equation,
these $q$-plane waves are the only traveling solutions of the form $f(kx-\omega t)$
admitted by (\ref{qoncer}). The effective velocity of these solutions
is $q$-dependent and given by,

\begin{equation}
c_ q = \frac{c}{\sqrt{1 + \frac{{\bar \delta}}{q} c^2}}.
\end{equation}


We have explored dissipation effects in the nonlinear Klein-Gordon
field theory recently introduced in \cite{NMT2011}. These effects
are described by a parameterized  evolution equation yielding
nonlinear versions of the celebrated telegraphists
equation. This equation incorporates as particular instances
various nonlinear evolution equations that are receiving
increasing attention recently, such as the power-law diffusion equation
(porous media equation) and the NRT nonlinear
Schroedinger and Klein-Gordon equations (the last one
corresponding to the NLKGD). The linear Klein-Gordon and
telegraphists equations are also recovered as particular limit cases.
The nonlinear telegraphists equation advanced here may be useful
for describing a variety of physical systems or processes such as,
for example, wave guides and electrical transmission lines with nonlinear
amplitude-depending dissipation, and nonlinear non-Poissonian dichotomous
diffusion processes.

\vskip 2cm

\noindent
{\bf Acknowledgments:}
Partial financial supports from Grant 401512/2014-2 of CNPq (Brazilian agency),
from FAPERJ (Brazilian agency), and from the John Templeton Foundation,
are acknowledged.


\end{document}